\documentclass{jfm}
\usepackage{graphicx}
\usepackage{epstopdf, epsfig}
\usepackage{authblk}

\shorttitle{Review Article}
\shortauthor{Naveen Balaji, Sujan Kumar, Seetharamu, Seetharam, Babu Rao, \& Rammohan}

\title{On Recent Developments in the Leading Edge Problem: Self-Similar Solutions to Momentum and Energy Equations of a Flat Plate}

\author[1]{U S Naveen Balaji}
\author[2]{Sujan Kumar S}
\author[3]{Dr. Kankanhally N Seetharamu}
\author[4]{Dr. T R Seetharam}
\author[5]{Babu Rao Ponangi}
\author[6]{Dr. Rammohan B}

\affil[1,2]{PES University,Bangalore -
560085, Karnataka, India, naveenandmetallica@gmail.com}
\affil[3,4]{Chair Professor, Department of Mechanical Engineering, PES University, Bangalore -
560085, Karnataka, India.}
\affil[5]{Assistant Professor, Department of Mechanical Engineering, PES University, Bangalore -
560085, Karnataka, India.}
\affil[6]{Associate Professor, Department of Mechanical Engineering, PES University, Bangalore -
560085, Karnataka, India.}

\begin{document}

\maketitle
 
\begin{abstract}
We provide an overview of the leading edge problem in this paper. We have used a self-similar function having a dependence on both the self-similar variable $\eta$ and Reynold's number $R$ to covert the momentum and energy equations into a fourth-order, non-linear partial differential equation (PDE) and a second-order, non-linear PDE respectively. Attempts have been made to solve the energy equation in a variety of ways, which include solving the PDE approximating the terms of the order $\mathcal{O}\left(R^{2}\right)$ and solving the PDE via the method of characteristics, but mostly being able to solve the energy PDE sans solving the momentum PDE. The complexities involved in solving the momentum PDE have been discussed and plausible approximate solutions have been given. The importance of boundary conditions and how they influence the solution to the energy PDE has been discussed. We have also shown how the energy PDE can be defined as a well-posed hyperbolic initial-boundary value problem in the leading edge. We conclude the paper by showing an approximate solution to the heat transfer coefficient and plot its characteristic behavior.
\end{abstract}

\begin{keywords}
Navier-Stokes Momentum PDE, Energy PDE, Self-Similar Solution Boundary-Layer Theory, Leading-edge Solution.
\end{keywords}

\section{The Leading Edge Problem: An Introduction}
\noindent The boundary layer solution as derived by \citet{Blasius} hold for large Reynold's numbers but becomes inaccurate at the leading edge \citet{vandyke}. Thus, it is necessary to obtain a consistent solution at the leading edge in order to better understand the origin of the Blasius solution. Boundary layer flow over a flat plate is a classical fluid mechanics problem that has received significant attention over the past years. A comprehensive collection of boundary layer theory is presented in \cite{Boundarytheory}. Prandtl's boundary theory provides a leading order solution to the asymptotic expansion of the Navier-Stokes equations for large Reynold's numbers. Several papers have tried to extend Blasius' solution to the leading edge. \cite{Alden} extended to higher-order terms and included the pressure distribution, which is neglected in the classical boundary layer theory. \cite{Kuo} used a technique proposed by \cite{Lighthill} by which the solution of an approximated non-linear equation can be extended in the neighbourhood of a singularity merely by straining the argument of the solution to improve Blasius' solution such that it's validity is extended up to the leading edge. Imai \cite{Imai} was able to derive a first-order solution to Prandtl's boundary layer theory and showed that the vorticity decays exponentially with distance from the plate. \cite{Avramenko} investigated self-similar boundary layers in nanofluids and found the solutions to be valid far downstream from the leading edge. A detailed description of the boundary layer theory along mathematical studies and their importance to the nonlinear theory of viscous flows is provided in \cite{Oleinik}.\\
\noindent In this paper, we build on the work done by \citet[]{Rao} and we study the laminar boundary layer over the leading edge of a flat semi-infinite plate. The governing equations and the boundary conditions are presented in section \ref{sec:goveq}, and the equations are expressed in terms of the streamfunction in section \ref{sec:stream}. The self-similar transforms are defined and the momentum PDE is expressed in the self-similar form in section \ref{sec:selfsimilar}. A self-similar energy equation is defined and presented as a mathematically well-posed problem in section \ref{sec:energy_eqn} and the difficulties associated with finding solutions and discussions pertaining to ambiguities in choice of boundary conditions are presented in section \ref{sec:energydifficult}, where we also find a general solution via the method of characteristics. An approximate solution to the energy PDE, neglecting terms of order $\mathcal{O}(R^{2})$ is presented and difficulties in obtaining boundary conditions is discussed in section \ref{sec:approxenergy}. All the work presented in sections \ref{sec:momsection} and \ref{sec:energy_eqn} is used to obtain an expression for the heat transfer coefficient and plot it's variation with the self-similar variable $\eta$ in section \ref{sec:heattransfer}.

\section{Problem Definition}\label{sec:momsection}
\subsection{Governing Equations and Boundary Conditions}\label{sec:goveq}
\noindent We consider an incompressible Newtonian fluid flowing with a uniform velocity $U$ into the leading edge of a semi-infinite flat plate. The fluid obeys the continuity and the momentum equations, given by
\\
\begin{equation}\label{conti}
    \frac{\partial u}{\partial x} + \frac{\partial v}{\partial y} = 0,
\end{equation}
\\
\begin{equation}\label{mom}
    \begin{array}{lr}
    \rho\left(u\frac{\partial u}{\partial x} +v\frac{\partial u}{\partial y} \right) = -\frac{\partial p}{\partial x} + \mu\left(\frac{\partial^{2} u}{\partial x^{2}} + \frac{\partial^{2} u}{\partial y^{2}} \right),\\\\
    
    \rho\left(u\frac{\partial v}{\partial x} +v\frac{\partial v}{\partial y} \right) = -\frac{\partial p}{\partial y} + \mu\left(\frac{\partial^{2} v}{\partial x^{2}} + \frac{\partial^{2} v}{\partial y^{2}} \right),
    \end{array}
\end{equation}
\\
where $u$ and $v$ are the $x$ and $y$ components of velocities respectively, $p$ is pressure, $\rho$ is density, and $\mu$ is the viscosity. The boundary conditions are that at the plate $y=0$, the fluid obeys no slip and no penetration, i.e.,
\\
\begin{equation}\label{atplate}
    u=0,\ v=0.
\end{equation}
\\
Far-field conditions dictate that as $y\rightarrow \infty$, the fluid obeys the free-stream conditions, i.e.,
\\
\begin{equation}\label{farfield}
    u\rightarrow U,\ v\rightarrow 0.
\end{equation}
\\
At the leading edge $x=0$, a uniform stream is imposed as follows
\\
\begin{equation}\label{leadingedge}
    u=U,\ v=0.
\end{equation}
\\
\subsection{Streamfunction}\label{sec:stream}
We now introduce the stream function $\psi$ in order to restate the momentum equations in terms of one single variable. We define
\\
\begin{equation}
    u=\frac{\partial \psi}{\partial y},\ v=-\frac{\partial \psi}{\partial x}.
\end{equation}
\\
The continuity equation in \ref{conti} is satisfied identically. The momentum equations as given in \ref{mom} become, with the subscript denoting differentiation,
\\
\begin{equation}
    \begin{array}{lr}
    \rho\left(\psi_{x}\psi_{xy} - \psi_{x}\psi_{yy} \right) = -p_{x} +\mu\left(\psi_{xxy} + \psi_{yyy} \right),\\\\
    
    \rho\left(-\psi_{y}\psi_{xx} + \psi_{x}\psi_{xy} \right) = -p_{y} -\mu\left(\psi_{xxx} + \psi_{xyy} \right).
    \end{array}
\end{equation}
\\
We now cross-differentiate to eliminate pressure and obtain
\\
\begin{equation}\label{pressureeliminator}
\rho\left(\psi_{y}\nabla^{2}\psi - \psi_{x}\nabla^{2}\psi \right) = \mu\nabla^{4}\psi,    
\end{equation}
\\
where $\nabla$ is a two-dimensional vector-operator, $\nabla = \left<\partial_{x},\partial_{y}\right>$. The boundary conditions can also be expressed in terms of the streamfunction. At the plate $y=0$, and \ref{atplate} yields
\\
\begin{equation}
    \psi_{y}=0,\ \psi_{x}=0.
\end{equation}
\\
Far from the plate as $y\rightarrow\infty$, \ref{farfield} gives
\\
\begin{equation}
    \psi_{y}\rightarrow U,\ \psi_{x}\rightarrow 0.
\end{equation}
\\
At the leading edge $x=0$, \ref{leadingedge} becomes
\\
\begin{equation}
    \psi_{y}=U,\ \psi_{x}=0.
\end{equation}
\\

\subsection{Self-Similar Transformation}\label{sec:selfsimilar}
\noindent At the leading edge, viscous forces dominate in \ref{pressureeliminator} and a balance between the viscous terms gives $y\sim x$. Hence, we define a self-similar variable
\\
\begin{equation}\label{selfsimvariable}
    \eta = \frac{y}{x},
\end{equation}
\\
and a self-similar streamfunction which depends on both the self-similar variable $\eta$ and the Reynold's number $R$ 
\\
\begin{equation}\label{selfsmfunc}
    f(\eta,R) = \frac{\psi}{Ux}.
\end{equation}
\\
Now, substituting \ref{selfsimvariable} and \ref{selfsmfunc} into \ref{pressureeliminator} transforms the equation into the following PDE
\\
\begin{equation}\label{momentumfourthPDE}
    \begin{array}{lr}
    \left(1+\eta^{2}\right)f_{\eta\eta\eta\eta} + 8\eta\left(1+\eta^{2}\right)f_{\eta\eta\eta} + 4\left(1+3\eta^{2} \right)f_{\eta\eta}\\\\
    +\left(2\eta ff_{\eta\eta} + \left(1+\eta^{2} \right)\left(ff_{\eta\eta}\right)_{\eta} -4\left(1+3\eta^{2}\right)f_{\eta\eta R} -4\eta\left(1+\eta^{2}\right)f_{\eta\eta\eta R}\right)R\\\\
    +\left(2\eta\left(ff_{R}f_{\eta\eta} - ff_{\eta\eta R}\right)- \left(1+\eta^{2} \right)\left(f_{\eta}f_{\eta\eta R} - f_{R}f_{\eta\eta\eta} \right) + 2\left(1+3\eta^{2}\right)f_{\eta\eta RR}\right)R^{2}\\\\
    +\left((2\eta\left(f_{\eta}f_{\eta RR}-f_{R}f_{\eta\eta R}\right) + ff_{\eta RR} -3f_{\eta}f_{RR} + 4f_{RRR})-4\eta f_{\eta RRR}\right)R^{3}\\\\
    +\left(f_{RRRR}+f_{R}f_{\eta RR} - f_{\eta}f_{RRR}\right)R^{4}=0,
    \end{array}
\end{equation}
\\
where $R=\rho Ux/\mu$ is the local Reynold's number. It is to be mentioned here that the expression given in \citet[]{Rao} is incorrect and the equation \ref{momentumfourthPDE} is the correct form of the self-similar momentum PDE. Note that \ref{momentumfourthPDE} is a fourth-order non-linear PDE and as we shall see, the solution to the PDE determines the characteristic behavior of the energy PDE and hence, the heat transfer coefficient. The boundary conditions can now be expressed in terms of the self-similar variables. We have discussed the ambiguities associated with obtaining boundary conditions in future sections since here, we have the self-similar streamfunction to depend on both the self-similar variable $\eta$ and $R$. Although we do know the behavior of the self-similar variable $\eta$ at the leading edge, at the plate, and far away from the plate we do not know how Reynold's number varies with changing $\eta$ and hence, determining the boundary conditions become cumbersome. We try to come-around this anomaly by assuming certain classes of Reynold's number models that relate the same to the self-similar variable $\eta$.

\section{The Energy Equation}\label{sec:energy_eqn}
\subsection{Problem Definition}
\noindent When $\psi=f(\eta,R)Ux$ and $T=T(\eta,R)$, where $R=Ux/\nu$, we obtain the following PDE
\\
\begin{equation}\label{energyeq}
    (1+\eta^{2})T_{\eta\eta}+2\eta T_{\eta}+R\left(Pr U-4\eta T_{\eta} \right)+R^{2}\left(T_{RR}-PrU\left(f_{\eta}f_{\eta R}-f_{R}f_{\eta \eta} \right) \right)=0,
\end{equation}
\\
where $Pr$ is the Prandtl number which we will set to unity for all our calculations. Notice that when $R\rightarrow 0$, \ref{energyeq} simplifies to the following
\begin{equation}
    2\eta T_{\eta}+\left(1+\eta^{2}\right)T_{\eta\eta}=0.
\end{equation}
 \\
Similarly, for $Pr\equiv 1$ and $R\in[0,\infty)$, \ref{energyeq} becomes
\\
\begin{equation}\label{Prenergyeq}
    (1+\eta^{2})T_{\eta\eta}+2\eta T_{\eta}+R\left(U-4\eta T_{\eta} \right)+R^{2}\left(T_{RR}-U\left(f_{\eta}f_{\eta R}-f_{R}f_{\eta \eta} \right) \right)=0.
\end{equation}
\\
We now write \ref{Prenergyeq} as follows
\\
\begin{equation}
    (1+\eta^{2})T_{\eta\eta}+R^{2}T_{RR}+2\eta(1-2R)T_{\eta}+G\left(f_{\eta},f_{R},f_{\eta R},f_{\eta\eta},R \right)=0,
\end{equation}
 \\
where
\\
\begin{equation}\label{energyPDE}
    G\left(f_{\eta},f_{R},f_{\eta R},f_{\eta\eta},R \right)=R^{2}U\left(f_{\eta}f_{\eta R}-f_{R}f_{\eta \eta} \right).
\end{equation}
\\
Now, comparing \ref{Prenergyeq} with the standard form of a second-order PDE written as
\\
\begin{equation}
    Au_{xx}+Bu_{xy}+Cu_{yy}+Du_{x}+Eu_{y}+Fu+G=0,
\end{equation}
\\
we observe that $B=0$, $A=(1+\eta^{2})$, and $C=R^{2}$. Consider the following cases
\\\\
i.\ When $R\rightarrow 0$, $B^{2}-AC=0$ and \ref{Prenergyeq} becomes a parabolic PDE.\\\\
ii.\ At the plate $y=0$ which implies that $\eta=0$ and hence, $B^{2}-AC>0$ and \ref{Prenergyeq} becomes a hyperbolic PDE.\\\\
iii.\ Far away from the plate $y\rightarrow \infty$ which implies that $\eta \rightarrow \infty$ and hence, $B^{2}-AC>0$ and \ref{Prenergyeq} becomes a hyperbolic PDE.\\\\
iv.\ At the leading edge $x=0$ which implies that $\eta\rightarrow \infty$ and hence, $B^{2}-AC>0$ and \ref{Prenergyeq} becomes a hyperbolic PDE.\\
\\\\
Define a non-linear operator $N$ as follows
\\
\begin{equation}
    N:=\frac{1}{R^{2}}\left(1+\eta^{2}\right)\frac{\partial^{2}}{\partial\eta^{2}}+2\eta(1-2R)\frac{\partial}{\partial \eta},
\end{equation}
\\
such that
\\
\begin{equation}\label{hyperenergy}
    N[T]+T_{RR}=H,
\end{equation}
\\
in $\mathcal{V}_{R}$ where $H=G\left(f_{\eta},f_{R},f_{\eta R},f_{\eta\eta},R \right)$ and this PDE is defined in the open set $\mathcal{V}\in \mathbb{R}^{2}$. We can now define a hyperbolic initial boundary value problem as follows
\\
\begin{equation}
    \begin{array}{lr}
    N[T]+T_{RR}=H\ \ \ in\ \mathcal{V}_{R},\\
    T=T_{\infty}\ \ \ \ \ \ \ \ \ \ \ \ \ \ \ on\ \partial\mathcal{V}\times [0,R],\\
    T=h,\ T_{R}=g\ \ \ \ \ \ on\ \mathcal{V}\times \{R=0\},
    \end{array}
\end{equation}
\\
where $\mathcal{V}$ is an open set in $\mathbb{R}^{2}$, $\mathcal{V}_{R}=\mathcal{V}\times(0,R]$ for a fixed Reynold's number $R>0$, $H:\mathcal{V}_{R}\rightarrow \mathbb{R}$ and $h,g:\mathcal{V}\rightarrow \mathbb{R}$ are given and $T:\bar{\mathcal{V}}_{R}\rightarrow \mathbb{R}$ is unknown, $T=T(\eta,R)$. Here $H:\mathcal{V}_{R}\rightarrow \mathbb{R}$ is given and $N$ denotes, for each Reynold’s number $R$ a second-order partial differential operator.

\subsection{Difficulty Associated with Solving the Energy Equation}\label{sec:energydifficult}
\noindent In order to solve for the heat transfer coefficient, we need to first find $f(\eta,R)$, i.e., the solution to the self-similar momentum equation. The momentum equation with the self-similar transformation is a fourth-order, non-linear equation which is subject to boundary conditions
\\
\begin{equation}\label{generalcondition1}
\begin{array}{lr}
    f_{\eta}=0,\\
    f+f_{R}R=0
\end{array}
\end{equation} 
\\
at the plate (i.e., $\eta=0$) and
\\
\begin{equation}\label{generalcondition2}
    \begin{array}{lr}
    f_{\eta}\rightarrow 1,\\
    f+f_{R}R\rightarrow \eta
    \end{array}
\end{equation}
\\
being the far-field conditions (i.e., at $\eta\rightarrow \infty$). These are the boundary conditions which are to be imposed in order to determine the constants of a general solution to the momentum PDE \ref{momentumfourthPDE}, in all future sections we refer to boundary conditions \ref{generalcondition1} and \ref{generalcondition2} as generalized boundary conditions. It is to be noted here that the momentum PDE \ref{momentumfourthPDE} is immensely difficult to solve and the existence of a closed-form solution is uncertain. The PDE was obtained from the steady-state Navier-Stokes momentum equations \ref{mom} after eliminating the pressure terms via cross-differentiation post the introduction of a streamfunction and by trying to find a solution of the form $f(\eta,R)$ we are trying to find a solution $\psi(x,y)$ to the pressure-less momentum PDE \ref{pressureeliminator} which in turn are the solutions $u(x,y)$ and $v(x,y)$ to the momentum equations \ref{mom} themselves. This turns out to be a simplified case of a more general problem-to prove that there exists a smooth solution $u(x,t)$ on $\mathbb{R}\times[0,\infty)$ to the unsteady Navier-Stokes equations. This is an open problem which has been listed as one of the millennium problems by the Clay Institute of Mathematics. It will be shown that although the conditions are satisfied at the plate, the far-field conditions are not. Although in this case, the characteristic curve of the heat-transfer coefficient is similar to, except for a singularity at $\eta\rightarrow 0$, the one obtained by our calculations.\\
\noindent
Now, since the leading edge is usually subjected to Reynold’s number of the order $10^{-3}$, we may ignore the terms of order $\mathcal{O}(R^{2})$ and above. This approximation yields the following PDE 
\\
\begin{equation}\label{momeqO1}
\begin{array}{lr}
    \left((1+\eta^{2})^{2}f_{\eta\eta\eta\eta}+8\eta(1+\eta^{2})f_{\eta\eta\eta} +4\eta(1+3\eta^{2})f_{\eta\eta}\right)\\
    + R\left(2\eta ff_{\eta\eta} +(1+\eta^{2})\left(ff_{\eta\eta}\right)_{\eta}-4(1+3\eta^{2})f_{\eta\eta R}-4\eta(1+\eta^{2})f_{\eta\eta\eta R}\right)=0.
    \end{array}
\end{equation}
\\
It is observed that the first part of the PDE, the part independent of $R$ can be solved by setting $u=f_{\eta\eta}$ in order to reduced the PDE to the following ODE
\\
\begin{equation}\label{momODE}
     \left((1+\eta^{2})^{2}u_{\eta\eta}+8\eta(1+\eta^{2})u_{\eta} +4\eta(1+3\eta^{2})u\right)=0,
\end{equation}
\\
whose solution reads
\\
\begin{equation}
    f(\eta) = \frac{2}{\pi}tan^{-1}(\eta) + \frac{2\eta}{\pi (1+\eta^{2})}.
\end{equation}
\\
The part of \ref{momeqO1} which involve terms of $\mathcal{O}(R)$ read
\\
\begin{equation}\label{momOR2}
    R\left(2\eta ff_{\eta\eta} +(1+\eta^{2})\left(ff_{\eta\eta}\right)_{\eta}-4(1+3\eta^{2})f_{\eta\eta R}-4\eta(1+\eta^{2})f_{\eta\eta\eta R}\right)=0.
\end{equation}
\\
This is a fourth-order non-linear PDE which is difficult to solve to obtain analytic solution. The difficulty associated with solving \ref{momOR2} can be attributed to the undifferentiated term $f(\eta,R)$ in $\left(ff_{\eta\eta}\right)_{\eta}$ since if we were to make the substitution $u=f_{\eta\eta}$ as made previously, we would obtain the following PDE
\\
\begin{equation}
\begin{array}{lr}
    R(2\eta u\left(\int{\int{u\ d\eta}}\ d\eta\right) +(1+\eta^{2})\left(u_{\eta}\left(\int{\int{u\ d\eta}}\ d\eta\right)+u\int{u\ d\eta}\right)\\
    -4(1+3\eta^{2})u_{R}-4\eta(1+\eta^{2})u_{\eta R})=0.
    \end{array}
\end{equation}
\\
which is a non-linear partial-integro differential equation and this complicates the problem. One possible method could be to remove the partial differentiation of $f$ by $R$ by assuming that the Reynold's number depends on the self-similar variable $\eta$ as $R=\alpha \eta^{\beta}$, where $\alpha$ and $\beta$ are real numbers. With this, \ref{momOR2} becomes a fourth-order non-linear PDE in $\eta$ and has the following form
\\
\begin{equation}
    R\left(2\eta ff_{\eta\eta} +(1+\eta^{2})\left(ff_{\eta\eta}\right)_{\eta}-\frac{4(1+3\eta^{2})}{\alpha\beta\eta^{\beta-1}}f_{\eta\eta\eta}-\frac{4(1+\eta^{2})}{\alpha\beta\eta^{\beta-2}}f_{\eta\eta\eta\eta}\right)=0,
\end{equation}
\\
but this resulting PDE is still complicated and difficult to solve. We here propose an alternative method where we notice that the PDE \ref{momeqO1} has only one differential in $R$ and thus, we set $f_{\eta\eta\eta R} = \gamma f_{\eta\eta\eta}$ and $f_{\eta\eta R} = \gamma f_{\eta\eta}$, where $\gamma$ is a positive constant. Now, we rewrite the second part of the PDE which is dependent on $R$ as follows
\\
\begin{equation}
    R\left(2\eta ff_{\eta\eta}(1+\eta^{2})\left(ff_{\eta\eta}\right)_{\eta} -4\gamma (1+3\eta^{2})f_{\eta\eta} - 4\eta\gamma(1+\eta^{2})f_{\eta\eta\eta} \right) = 0,
\end{equation}
\\
which can now be split into
\\
\begin{equation}\label{PDE1}
    (1+\eta^{2})\left(ff_{\eta\eta} \right)_{\eta}-4\gamma(1+3\eta^{2})f_{\eta\eta}=\psi(R),
\end{equation}
\\
and
\\
\begin{equation}\label{PDE2}
     2\eta ff_{\eta\eta} - 4\eta\gamma(1+e\eta^{2})f_{\eta\eta\eta}=\phi(R),
\end{equation}
\\
where $\phi(R)$ and $\psi(R)$ are arbitrary functions of the Reynold's number. We now substitute the expression for $f_{\eta\eta\eta}$ which can be obtained from \ref{PDE2} into \ref{PDE1} to obtain the following
\\
\begin{equation}
    \left((1+\eta^{2})f_{\eta}-4\gamma(1+3\eta^{2})\right)f_{\eta\eta}+\frac{1}{2\gamma}f^{2}f_{\eta\eta}- \frac{\psi(R)}{4\eta\gamma}f=\phi(R).
\end{equation}
\\
Now, setting $\phi(R)=0,\ \psi(R)=0,$ and $\gamma=1$, the PDE simplifies to the following
\\
\begin{equation}
    f_{\eta\eta}\left((1+\eta^{2})f_{\eta}-4(1+3\eta^{2})+\frac{1}{2}f^{2}\right) = 0,
\end{equation}
\\
which has the following solution
\\
\begin{equation}\label{momeq01}
    \left(f(\eta,R)\right)_{\mathcal{O}(R)} = \frac{  4\left(4 + 5\eta^{2}+3\eta^{4} + 
   3\eta (1 + \eta^{2})^{2} tan^{-1}(\eta) - 
   16\eta (1 + \eta^{2})^{2} C_{1}\right)}{\eta(5+3\eta^{2})+2(1+\eta^{2})^{2}tan^{-1}(\eta)-16(1+\eta^{2})^{2}C_{1}}.
\end{equation}
\\
To fix the constant $C_{1}$ we make use of the boundary condition at the plate, i.e., \ref{generalcondition1}. We find that the constant $C_{1}=\pm\left( 1/2\sqrt{2}\right)$. Now, the solution to the momentum PDE upto order $\mathcal{O}(R)$ would be a linear combination of the solutions of the first part of \ref{momeqO1} which is independent of $R$, i.e., $f(\eta)$ and the second part of \ref{momeqO1} which depends on $R$ upto order $\mathcal{O}(R)$, i.e., $\left(f(\eta,R) \right)_{\mathcal{O}(R)}$. Thus, the final solution for positive constants $\lambda$ and $\chi$ read
\\
\begin{equation}\label{momsol}
\begin{array}{lr}
    f(\eta,R)= \lambda f(\eta) + R\chi\left(f(\eta,R) \right)_{\mathcal{O}(R)}\\
    \ \ \ \ \ \ \  =\lambda\left(\frac{2}{\pi}tan^{-1}(\eta) +\frac{2\eta}{\pi(1+\eta^{2})}\right) + R\chi\frac{  4\left(4 + 5\eta^{2}+3\eta^{4}+3\eta (1 + \eta^{2})^{2} tan^{-1}(\eta) - 
   16\eta(1+\eta^{2})^{2}\left(\pm\frac{1}{2\sqrt{2}}\right)\right)}{\eta(5+3\eta^{2})+2(1+\eta^{2})^{2}tan^{-1}(\eta)-16(1+\eta^{2})^{2}\left(\pm\frac{1}{2\sqrt{2}}\right)}.
\end{array}
\end{equation}
\\
Note that this is only an approximate solution to the momentum PDE where we have neglected terms of order $\mathcal{O}(R^{2})$. We can now use this result to solve the energy equation without having to compromise terms of order $\mathcal{O}(R^{2})$ although this too will only be an approximate solution due to the compromise made in the momentum PDE.
\\
\subsection{Approximate Solution to the Energy Equation- Method of Characteristics}
\noindent The energy PDE \ref{Prenergyeq} can be rewritten using the result obtained in \ref{momsol} and by doing so we attempt to solve the energy PDE by the method of characteristics. If the PDE $N[T]+T_{RR}=H$ is hyperbolic in the domain $\mathcal{V}$, it implies that $B^{2}-AC>0$ at each point of $\mathcal{V}$. We now choose $(\eta, R)\rightarrow \left(\omega(\eta,R),\gamma(\eta,R) \right)$such that 
\\
\begin{equation}
    A(\omega,\gamma) = a\omega^{2}_{\eta} + b\omega_{\eta}\omega_{R} +c\omega_{R}^{2}=0,\ C(\omega,\gamma) = a\gamma_{\eta}^{2}+b\gamma_{\eta}\gamma_{R}+c\gamma_{R}^{2}, 
\end{equation}
\\
which can be expressed as the following product
\\
\begin{equation}
    a\left[\omega_{\eta}-\frac{-b-\sqrt{b-ac}}{a}\omega_{R}\right]\cdot\left[\omega_{\eta}-\frac{-b+\sqrt{b-ac}}{a}\omega_{R}\right]=0,
\end{equation}
\\
and hence, we solve the following linear equations
\\
\begin{equation}
    \omega_{\eta} - \mu_{1}\omega_{R}=0,\ \ \omega_{\eta} - \mu_{2}\omega_{R} = 0,
\end{equation}
\\
where $\mu_{1}=-\frac{-b-\sqrt{b-ac}}{a}=-i\frac{R}{\sqrt{1+\eta^{2}}}$ and $\mu_{2} = \frac{b-\sqrt{b-ac}}{a}=i\frac{R}{\sqrt{1+\eta^{2}}}$. We now find that 
\\
\begin{equation}
    \omega = R\left(e^{-i\ sinh^{-1}(\eta)} \right),\ \gamma = R\left(e^{i\ sinh^{-1}(\eta)} \right),
\end{equation}
\\
and we can write the $T(\eta,R)=v\left(\omega(\eta,R),\gamma(\eta,R) \right)$. The energy PDE takes the following canonical form
\\
\begin{equation}
\begin{array}{lr}
    \omega^{2}v_{\omega\omega}-\gamma^{2}v_{\gamma\gamma} +\left(K(\omega,\gamma)+\sqrt{\omega\gamma} \right)(\omega v_{\omega} + \gamma v_{\gamma}) + W(\omega,\gamma)(\gamma v_{\gamma} - \omega v_{\omega})\\
    + G(f_{\eta},f_{R},f_{\eta R},f_{\eta\eta},R)=0,
\end{array}
\end{equation}
\\
where
\\
\begin{equation}
\begin{array}{lr}
    K(\omega,\gamma) = \frac{\sqrt{1+sinh^{2}\left(ln\left(\frac{\gamma}{\omega}\right)^{\frac{1}{2i}} \right)}-sinh\left(ln\left(\frac{\gamma}{\omega}\right)^{\frac{1}{2i}} \right)}{\sqrt{\omega \gamma}\sqrt{1+sinh^{2}\left(ln\left(\frac{\gamma}{\omega}\right)^{\frac{1}{2i}} \right)}}\\
    W(\omega,\gamma) = \frac{2sinh\left(ln\left(\frac{\gamma}{\omega}\right)^{\frac{1}{2i}} \right)(1-2\sqrt{\omega \gamma})}{\sqrt{1+sinh^{2}\left(ln\left(\frac{\gamma}{\omega}\right)^{\frac{1}{2i}} \right)}}.
\end{array}
\end{equation}
\\
With the characteristics found we can write the general solution to the hyperbolic PDE as follows
\\
\begin{equation}
  T(\eta,R) = F\left[ R\left(e^{-i\ sinh^{-1}(\eta)} \right)\right] +B\left[ R\left(e^{i\ sinh^{-1}(\eta)} \right)\right] + G(f_{\eta},f_{R},f_{\eta R},f_{\eta\eta},R), 
\end{equation}
\\
where $F$ and $B$ are functions of the characteristics $\omega$ and $\gamma$ respectively and $G(f_{\eta},f_{R},f_{\eta R},f_{\eta\eta},R) = U(f_{\eta}f_{\eta R}-f_{R}f_{\eta\eta})$. Using Euler's formula, we have the real part of the solution to be
\\
\begin{equation}
     T(\eta,R) = Re\left(F\left[ R\left(e^{-i\ sinh^{-1}(\eta)} \right)\right]\right) +Re\left(B\left[ R\left(e^{i\ sinh^{-1}(\eta)} \right)\right]\right) + G(f_{\eta},f_{R},f_{\eta R},f_{\eta\eta},R).
\end{equation}
\\

\subsection{Approximate Solution to the Energy Equation}\label{sec:approxenergy}
\noindent We here present a solution to the energy PDE without considering terms of order $\mathcal{O}(R^{2})$. Consider the energy PDE \ref{energyPDE} with this mentioned approximation
\\
\begin{equation}
    R\left(U-4\eta T_{\eta}\right)+2\eta T_{\eta} +\left(1+\eta^{2} \right)T_{\eta\eta} = 0.
\end{equation}
\\
We find a general solution of the following form
\\
\begin{equation}
    T(\eta,R) = \int_{1}^{\eta}{\left(-UR\  _{2}F_{1}\left[\frac{1}{2},2R,\frac{3}{2},-K_{1}^{2}\right]K_{1}\left(1+K_{1}^{2}\right)^{2R-1}+\left(1+K_{1}^{2}\right)^{2R-1}C_{1}\right)}+C_{2},
\end{equation}
\\
where $_{2}F_{1}$ is the hypergeometric function and $K_{1}$ is the modified Bessel function of the second kind. This integral is immensely complicated to solve and hence we expand the hypergeometric series around $K_{1}=0$ to obtain
\\
\begin{equation}
    _{2}F_{1}\left[\frac{1}{2},2R,\frac{3}{2},-K_{1}^{2}\right] = 1-\frac{2}{3}RK_{1}^{2}+\frac{1}{5}\left(R+2R^{2}\right)K_{1}^{4} +\mathcal{O}\left((K_{1})^{6}\right).
\end{equation}
\\
Now, integrating this expansion, the solution reads
\\
\begin{equation}
\begin{array}{lr}
    T(\eta,R) \approx C_{2} + C_{1}\eta - \frac{1}{2}(UR)\eta^{2} + \frac{1}{3}(2R-1)C_{1}\eta^{3} +\frac{1}{12}U(3-4R)R\eta^{4}\\\\
    +\frac{1}{5}(R-1)(2R-1)C_{1}\eta^{5} - \frac{1}{90}\left(UR(4R-5)(4R-3)\right)\eta^{6}\\\\ +\frac{1}{21}(R-1)(2R-3)(2R-1)C_{1}\eta^{7}   
    \end{array}
\end{equation}
\\
Before the fluid flows past the plate, there is no heat transfer and hence the temperature is unaffected. Thus, we demand that $T(\eta=0,R) = T_{\infty}$ and hence, we observe that $C_{2}=T_{\infty}$ which turns the solution into the form
\\
\begin{equation}
\begin{array}{lr}
T(\eta,R) = -\frac{1}{180}RU\eta^{2}\left(90+15(4R-3)\eta^{2}+ (30-64R+32R^{2})\eta^{4} \right) + T_{\infty}\\\\
+ \frac{1}{105}\eta\left(105 +35(2R-1)\eta^{2} + 21(1-3R +2R^{2})\eta^{4} + 5(-3+11R-12R^{2}+4R^{3})\eta^{6}\right)C_{1} 
\end{array}
\end{equation}
\\
Now, $R\rightarrow 0$ can imply two things-either $x=0$ and hence $\eta\rightarrow \infty$ (which is the leading edge condition) or the velocity of the fluid is very small. If the plate is being maintained at a uniform temperature then the temperature at all points of the plate would not fluctuate when there is no fluid available for heat transfer. Thus, we obtain a condition $T(\eta\rightarrow\infty,R\rightarrow0) = T_{w}$ which is valid only in the leading edge. Note however that $\eta\rightarrow\infty$ is also a condition when we are far away from the plate, i.e., when $y\rightarrow\infty$ and at this position even when there is no fluid flowing the temperature is that of the free-stream. Thus, the far-field condition reads $T(\eta\rightarrow\infty,R\rightarrow 0) = T_{\infty}$. When there is a fluid flowing with some arbitrary velocity, then far away from the plate there would be gradients of temperature as a result of heat transfer between the plate and the flowing fluid. Here, we don't make any assumption regarding the characteristic of the gradients since we don't know the behavior of temperature with viscosity and a similar argument can be made at the plate when $y\rightarrow 0$ for arbitrary Reynold's number. Thus, in order to establish the boundary conditions at the plate and far-field for arbitrary Reynold's numbers, we are to consider temperature-viscosity models, each of which depend upon the type of fluid begin used. Assuming that we have an appropriate model which provides a relationship between the temperature and viscosity, there still exists an additional constraint-relationship between the Reynold's number and the self-similar variable. In order to overcome this we are to yet again resort to assuming Reynold's number models as done in the previous sections.\\

\noindent To demonstrate this let us assume the following dependence of temperature on it's variables $\eta$ and $R$
\begin{equation}
    T(\eta,R) = \alpha tan^{-1}(\eta^{\beta}) + \gamma tan^{-1}(R^{\epsilon}),
\end{equation}
where $\alpha,\beta,\gamma,$ and $\epsilon$ are positive constants. In this model, at the plate
\\
\begin{equation}
\begin{array}{lr}
    T(\eta\rightarrow 0, R)=\gamma tan^{-1}(R^{\epsilon}),\\ T(\eta\rightarrow 0, R\rightarrow 0) = 0\ \& \\
    T(\eta\rightarrow 0,R\rightarrow\infty) = \gamma\frac{\pi}{2}.
\end{array}
\end{equation}
At the leading edge, 
\\
\begin{equation}
\begin{array}{lr}
    T(\eta\rightarrow \infty, R)=\alpha \frac{\pi}{2}+\gamma tan^{-1}(R^{\epsilon}),\\
    T(\eta\rightarrow\infty, R\rightarrow 0) = 0\ \& \\
    T(\eta\rightarrow \infty,R\rightarrow\infty) = \alpha\frac{\pi}{2} + \gamma\frac{\pi}{2}.
\end{array}
\end{equation}
\\
As we had deduced earlier, far away from the plate, $T(\eta\rightarrow\infty, R\rightarrow 0) = T_{\infty}$ but in order to find $T(\eta\rightarrow \infty , R)$ and $T(\eta\rightarrow \infty,R\rightarrow\infty)$ we need to consider temperature-viscosity models such as the Seeton or the Wright model. Similarly, when there is heat transfer between the fluid and the plate, temperature gradients would develop which would be of the following form
\\
\begin{equation}
    \left(\frac{dT}{dy} \right)_{y=0} = -\frac{h(T_{w}-T_{\infty})}{k},
\end{equation}
\\
and since $T_{y} = xT_{\eta}$ for a fixed value of $x$, we have $T_{\eta}(\eta=0,R) = -\frac{hx(T_{w}-T_{\infty})}{k}$. Notice now that the value of $x$ can be fixed by our choice of $R$ and thus, at the plate we obtain the following variations of the temperature gradients for different choices of Reynold's number
\\
\begin{equation}
    \begin{array}{lr}
    T_{\eta}(\eta=0,R) = -\frac{hx(T_{w}-T_{\infty})}{k},\\
    T_{\eta}(\eta=0,R\rightarrow 0) = 0,\ \& \\
    T_{\eta}(\eta=0,R\rightarrow\infty) \rightarrow -\infty.
    \end{array}
\end{equation}
\\
There would also exist a gradient with respect to the Reynold's number but this would be determined with the choice of the temperature-viscosity model chosen. For instance, if we choose that Seeton model, we have the following relation between the temperature and the kinematic viscosity
\\
\begin{equation}
    A+B\ ln(T) = ln\left(ln\left(\nu + 0.1 + e^{-\nu}K_{0}\left(\nu + 1.244067 \right) \right) \right),
\end{equation}
\\
where $A$ and $B$ are liquid-specific values and $K_{0}$ is the zero-order modified Bessel function of the second kind. We then obtain the following expression for $T$ as a function of $\nu$
\\
\begin{equation}
    T(\nu) = e^{-\frac{A}{B}} \left(ln\left(\nu + e^{-\nu}K_{0}\left(\nu + 1.244067 \right) \right) \right)^{\frac{1}{B}}.
\end{equation}
\\
Since the Reynold's number depends on both the kinematic viscosity and the self-similar variable we can probe for separable solutions of the form
\\
\begin{equation}
    \bar{T}(R) = \Omega(\eta)T(\nu).
\end{equation}
\\
and hence for the separable solution $T(\eta,R)$ which reads
\\
\begin{equation}
    T(\eta,R) = \Phi(\lambda)\Lambda(R),
\end{equation}
\\
where $\Lambda(r) = \lambda \bar{T}(R)$ for some positive constant $\lambda$.

\section{Expression for the Heat Transfer Coefficient}\label{sec:heattransfer}
\noindent Let us define the non-dimensional temperature as follows
\\
\begin{equation}\label{nontem}
    \theta = \frac{T(\eta,R) - T_{w}}{T_{\infty} - T_{w}},
\end{equation}
\\
where $T_{w}$ s the wall temperature and $T_{\infty}$ is the free-stream temperature, such that 
\\
\begin{equation}
    h = -k\left(\frac{\partial \theta}{\partial y}\right)_{y=0}.
\end{equation}
\\
At the plate, $y=0$ which implies that $\eta = 0$  and hence the non-dimensional temperature \ref{nontem} takes the following form
\\
\begin{equation}
    \theta(\eta,R) = \frac{F\left[ R\left(e^{-i\ sinh^{-1}(\eta)} \right)\right] +B\left[ R\left(e^{i\ sinh^{-1}(\eta)} \right)\right] + U(f_{\eta}f_{\eta R}-f_{R}f_{\eta\eta})-T_{w}}{T_{\infty} - T_{w}}.
\end{equation}
\\
Hence, 
\\
\begin{equation}
    \frac{\partial \theta}{\partial y} = \frac{\partial \theta}{\partial \eta}\frac{\partial \eta}{\partial y} = \frac{\theta_{\eta}}{x},
\end{equation}
\\
and we finally obtain the heat transfer coefficient to be the following
\\
\begin{equation}\label{heatrans}
    h(x) = -\frac{kU}{x(T_{\infty}-T_{w})}(f_{\eta}f_{\eta R}-f_{R}f_{\eta\eta})_{\eta=0},
\end{equation}
\\
where $k$ is the thermal conductivity of the fluid flowing past the plate. We observe now that finding the solution to the momentum PDE, i.e., $f(\eta,R)$ is essential in order to solve for the heat transfer coefficient. Since we have derived only an approximate solution for the momentum PDE (see equation \ref{momsol}), the solution to the heat transfer coefficient will also be approximate. Firstly, we notice that \ref{momsol} takes on an indeterminate form  at $\eta=\infty$ which is at the leading edge and yields a null result at $\eta=0$ which is at the plate. Thus, we study how the heat transfer coefficient varies with the self-similar transformation $\eta$ at the leading edge, i.e., with $x$ in the order of $10^{-3}$ units. We now compute the term enclosed in the brackets by setting the linear combination coefficients to unity (i.e., $\chi,\lambda\equiv1$) for simplicity. We choose the plate material to be made up of Aluminium, assume air to flow at a velocity $U=10\ m/s$, temperature of the plate to be $T_{w}=100^{0}C$ and the ambient temperature to be $T_{\infty} = 27^{0}C$.\\
\noindent 
\begin{figure}
  \centering{\includegraphics[width=10cm]{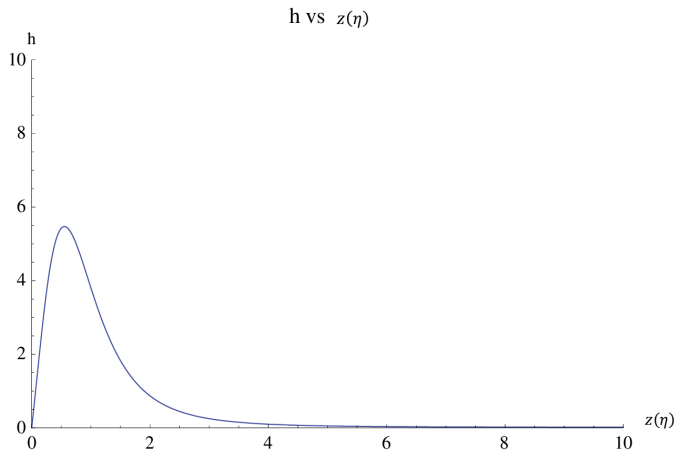}}
  \caption{This plot shows the variation of the heat transfer coefficient with the combination of the differential quantities contained in $z(\eta)$ and with $C_{1}=-\left(1/2\sqrt{2}\right)$ and $R=0.001$.}
\label{fig:hvseta}
\end{figure}
\\
Since we do not know the exact value of $\gamma$, we cannot obtain numerical results for $h(x)$. Hence, we plot $h(x)$ vs $z(\eta)$, where $z(\eta)=\left(f_{\eta}f_{\eta R} - f_{R}f_{\eta\eta}\right)$ to observe the characteristic behavior of the heat transfer coefficient. This is shown in figure \ref{fig:hvseta}.
\\
\section{Conclusion}
\noindent In this work, we have provided an overview of the leading edge problem and have studied the incompressible laminar flow over the leading edge of a semi-infinite flat plate. We have used a self-similar function, which has dependence on both the self-similar variable and the Reynold's number, to transform the energy and the momentum equations showing them to be second-order non-linear and fourth-order non-linear PDE respectively. The energy PDE has been formulated as a well-posed hyperbolic initial value problem on the leading edge and the difficulties associated with solving the same has been discussed. An approximate solution to the energy PDE is presented in which we have used the method of characteristics to solve the same and another approximate solution in which we do not consider terms of order $\mathcal{O}\left(R^{2} \right)$ has also been discussed.
The work also reveals deep connections between the momentum and the energy PDE since the latter cannot be solved for without a solution to the former. We have also discussed the importance of boundary conditions and shown how they influence the solution to the energy PDE. We have been able to find an expression for the heat transfer coefficient and have obtained its characteristic behavior at the leading edge by plotting its variation with the self-similar variable. We plan to solve the boundary layer numerically by a finite-volume method using CFD packages and also cross-verify the characteritic behavior of the heat-transfer coefficient via Monte Carlo simulations in future works.
\\
\section*{Acknowledgements}
\noindent The authors (Naveen Balaji and Sujan Kumar) would like to thank Dr. K N Seetharamu for introducing them to the leading edge problem, for his guidance and constant motivation, Dr. T R Seetharam for his deep insights and discussions, Mr. Babu Rao Ponangi for his support and Dr. Rammohan B for his deep insights and computational support. One of the authors (NB) would like to thank Dr. Vittal Rao of IISC, Bangalore for the multiple discussion sessions and Dr. Rodolfo Rosales of MIT, Boston for his deep insights and suggestions of literature. 

\bibliographystyle{jfm}
\bibliography{jfm-instructions}

\end{document}